\documentclass[fleqn,twoside]{article}
\usepackage{espcrc2}

\usepackage{graphicx}
\usepackage[figuresright]{rotating}

\newcommand{\numu}{\mbox{$\nu_{\mu}$}}

\newcommand{\stw}{\mbox{$\sin^2\theta_W$}}

\newcommand{\AmS}{{\protect\the\textfont2
  A\kern-.1667em\lower.5ex\hbox{M}\kern-.125emS}}

\title{Low Energy Neutrino Cross Sections: Comparison of Various Monte Carlo 
Predictions to Experimental Data}

\author{G.P. Zeller\address[Columbia]{Columbia University, 
Department of Physics, New York, NY 10027}}
       
\begin{document}

\begin{abstract}
Charged current (CC) and neutral current (NC) low energy neutrino cross 
section predictions from a variety of Monte Carlo generators in present use 
are compared against existing experimental data. Comparisons are made
to experimental data on quasi-elastic, resonant and coherent single pion 
production, multiple pion production, single kaon production, and total 
inclusive cross sections, and are restricted to the case of $\numu$ scattering
off free nucleons.

\vspace{1pc}
\end{abstract}

\maketitle

\section{Introduction}

\begin{figure}[h]
  \includegraphics[angle=90,bbllx=113bp,bblly=67bp,bburx=506bp,bbury=741bp,width=18.5pc]{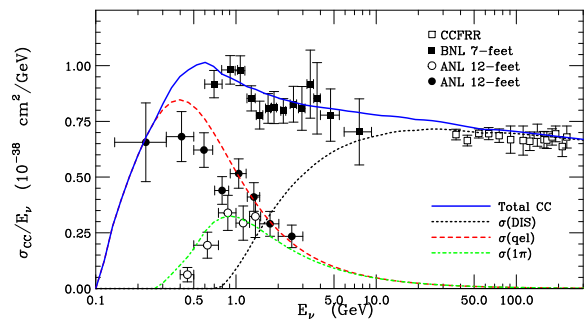}
  \vspace{-0.6in}
  \caption{Charged current neutrino cross sections as a function of energy 
          (in GeV). Shown are the contributions from quasi-elastic (dashed), 
          single pion (dot-dash) and deep inelastic scattering (dotted) 
          processes. Figure from Reference~\cite{lipari}.}
  \label{fig:lipari}
\end{figure}

Present atmospheric and accelerator based neutrino oscillation experiments 
operate at low neutrino energies ($E_\nu\sim1$ GeV) to access the relevant 
regions of oscillation parameter space. As such, they require precise knowledge
of the cross sections for neutrino-nucleon interactions in the sub-to-few
GeV range. At these energies, neutrino interactions are predominantly 
quasi-elastic (QE) or single pion production processes
(Figure~\ref{fig:lipari}), which historically have not been as well studied 
as deep inelastic scattering (DIS) reactions which dominate at higher 
energies. 

\vspace{-0.35in}

\vspace{-0.1in}
\noindent
Data on quasi-elastic scattering and single pion production come mainly 
from bubble chamber, spark chamber, and emulsion experiments that ran decades 
ago. Despite relatively poor statistics and large neutrino flux uncertainties,
they provide an important and necessary constraint on Monte Carlo models
in present use. Recent neutrino experiments employ a variety of Monte Carlo
generators to model low energy neutrino interaction cross sections. Many share
common theoretical inputs such as Llewellyn Smith free nucleon QE cross 
sections~\cite{ll-smith}, Rein and Sehgal-based resonance 
production~\cite{rein-sehgal}, along with standard DIS formulas and parton 
distribution functions. Yet the generators can differ substantially in how 
they implement Fermi gas models, combine resonance and DIS regions, and treat 
nuclear and final state effects~\cite{lipari}. This work is an attempt to 
compare several Monte Carlo generators to the existing body of low energy 
neutrino cross section data. In particular, three simulations are considered: 
v2 NUANCE~\cite{nuance}, NEUGEN~\cite{neugen}, and NUX~\cite{nux}. For the 
purpose of comparison, the Monte Carlo cross section predictions in each case
have been generated under the same set of parameter 
assumptions~\cite{qe-inputs}, namely: $m_A=1.032$ GeV, $m_V=0.84$ GeV, 
$g_A=-1.25$, and $\stw=0.233$ where applicable. While only free nucleon cross 
section comparisons are presented here, detailed comparisons of generated 
Monte Carlo event kinematics are provided in~\cite{dave-kinematics}.

\section{Quasi-Elastic Scattering}

At low energies, CC neutrino hadron interactions are 
predominantly quasi-elastic:

\vspace{-0.2in}
\begin{eqnarray*}
   \numu \, n \rightarrow \mu^- \, p
\end{eqnarray*}

\noindent
In predicting the cross section for quasi-elastic scattering off free 
nucleons, most Monte Carlos commonly employ the Llewellyn Smith 
formalism~\cite{ll-smith} in which the QE differential cross section takes 
the form:

\begin{eqnarray}
\label{eqn:xsectqe}
  \frac{d\sigma}{dQ^2} = \frac{G_F^2 M^2}{8\pi E_\nu^2} 
                        \left[ A \mp \frac{(s-u)}{M^2}B + 
                               \frac{(s-u)^2}{M^4} C \right] 
\end{eqnarray}

\noindent
where $(+)-$ refers to (anti)neutrino scattering, $G_F$ is the Fermi 
coupling constant, $Q^2$ is the squared four-momentum transfer ($Q^2=-q^2>0$), 
$M$ is the nucleon mass, $E_\nu$ is the incident neutrino energy, 
and $(s - u) = 4ME_\nu - Q^2 - m^2$. The factors $A$, $B$, and~$C$ are 
functions of the two vector form factors $F_1$ and $F_2$, the axial 
vector form factor $F_A$, and the pseudoscalar form factor~$F_P$:

\begin{eqnarray}
  A &=& {(m^2+Q^2)\over M^2}\left[ \left(1 + \tau \right)F_A^2 
        - \left( 1 - \tau \right)F_1^2 \right. \nonumber \\ 
    && \left. + \tau 
        \left( 1 - \tau \right) F_2^2 
        + 4 \tau F_1 F_2 \right. \nonumber \\
    && \left.\mbox{} - {m^2\over 4M^2} \left( (F_1 + F_2)^2 + 
        (F_A + 2F_P)^2 \right. \right. \nonumber \\
    && \left. \left. \hspace{0.9in} 
       - \left( {Q^2\over M^2} +4 \right) F_P^2 \right)\right] \\
 B &=& \frac{Q^2}{M^2} F_A (F_1 + F_2)  \\
\label{eqn:xsectabc}
 C &=& {1\over 4}\left(F_A^2 + F_1^2 + \tau F_2^2 \right) 
\end{eqnarray}

\noindent
where $\tau=Q^2/4M^2$ and $m$ is the muon mass. Monte Carlos commonly 
assume a dipole form for the factors $F_1$, $F_2$, $F_A$, and $F_P$:

\begin{eqnarray}
\label{eqn:form-factors}
   F_1(Q^2)  &=&  {{\displaystyle{ 1 + \tau \left( 1+\mu_p - \mu_n \right) 
            \over {\left(\displaystyle{1 + \tau} \right)
            \left(\displaystyle{1 + {{Q^2}\over{m_V^2} }}
            \right)^2}}}} \\
   F_2(Q^2)  &=&  {{{(\mu_p - \mu_{n})}} \over 
                  {{\left(\displaystyle{1 + \tau } \right)}
             \left(\displaystyle{1 + {{Q^2}\over{m_V^2}}} \right)^2}} \\
    F_A(Q^2) &=&  {{g_{A}} \over {\left(\displaystyle{1 + {{Q^2}\over{m_A^2} }}
             \right)^2}};  \hspace{0.2in} g_A=-1.25 \\
    F_P(Q^2) &=& {{{2 M^2} \over {m_{\pi}^2 + Q^2}} F_A(Q^2)}
\end{eqnarray}

\noindent
where $m_\pi$ is the pion mass, $\mu_p=1.793 \, \mu_N$ and 
$\mu_n=-1.913 \, \mu_N$ are the proton and neutron anomalous magnetic moments,
respectively, and the parameters $m_V$, $m_A$, and $g_A$ are empirical 
inputs~\cite{qe-inputs}. Departures from this dipole approximation which 
better fit electron scattering data have recently been explored and have 
few-$\%$ effects on the shape of the predicted cross section~\cite{bodek}.

\begin{figure}[h]
  \includegraphics[angle=0,width=16.6pc]{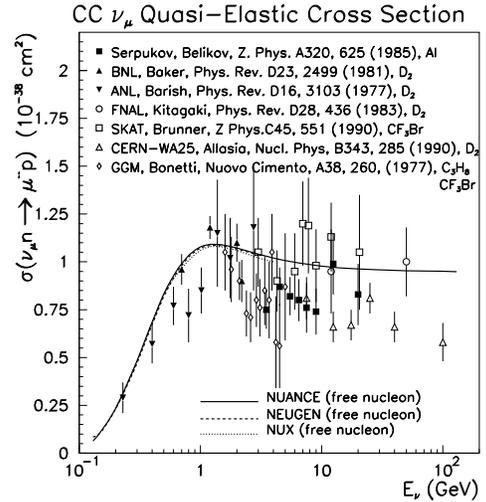}
  \vspace{-0.4in}
  \caption{QE data compared to various Monte Carlo calculations assuming
           a dipole form for the vector and axial vector form factors with
           $m_V=0.84$ GeV, $m_A=1.032$ GeV, and $g_A=-1.25$.}
  \label{fig:qe-data}
\end{figure}

Over the years, quasi-elastic processes have been studied extensively
in bubble chamber experiments at ANL, BNL, CERN, FNAL, 
and Serpukhov. The bulk of this data came from light targets and had 
limited precision due to large neutrino flux uncertainties. 
Figure~\ref{fig:qe-data} compares the various Monte Carlo predictions to
this collection of QE cross section data. It is no surprise that given 
the same underlying model and input parameters, the various Monte Carlo 
predictions agree up to their numerical precision; however, note that
there is a large spread in the available experimental data.

\vspace{-0.4in}
\clearpage

\section{NC Elastic Scattering}

Neutrinos can also elastically scatter from both protons and neutrons
in the target material:

\vspace{-0.2in}
\begin{eqnarray*}
   \numu \, p \rightarrow \numu \, p \\
   \numu \, n \rightarrow \numu \, n 
\end{eqnarray*}

\noindent
Equations~\ref{eqn:xsectqe}-\ref{eqn:xsectabc} still apply in describing the
neutral current elastic scattering cross section with the exception that in
this case the form factors include additional coupling factors and a 
contribution from strange quarks:

\vspace{-0.1in}
\begin{eqnarray}
 F_1(Q^2) &=& \left(\frac{1}{2}-\sin^2\theta_W\right)
          \left[\frac{\tau_3(1+\tau\,(1+\mu_p-\mu_n))}
                     {(1+\tau)\left(1+Q^2/m_V^2\right)^2}\right]
\nonumber \\ 
&& \hspace{-0.4in} -\sin^2\theta_W \left[\frac{1+\tau\,(1+\mu_p+\mu_n)}
                       {(1+\tau)\left(1+Q^2/m_V^2\right)^2}\right] 
  - \frac{F_1^s(Q^2)}{2} \nonumber \\
 F_2(Q^2) &=& \left(\frac{1}{2}-\sin^2\theta_W\right){\tau_3\,{(\mu_p-\mu_n)} 
      \over {(1+\tau) \left(\displaystyle{1 + {{Q^2}\over{m_V^2} }}\right)^2}} 
\nonumber \\
      && \hspace{-0.4in} -\sin^2\theta_W {{\mu_p+ \mu_n} 
      \over {(1+\tau) \left(\displaystyle{1 + {{Q^2}\over{m_V^2} }}\right)^2}} 
      - \frac{F_2^s(Q^2)}{2} \nonumber \\
 F_A(Q^2) &=&   {{g_{A} \: \tau_3} 
             \over {2 \left(\displaystyle{1 + {{Q^2}\over{m_A^2} }}\right)^2}} 
             + \frac{F_A^s(Q^2)}{2} \nonumber
\end{eqnarray}

\noindent
where $g_A=-1.25$ and $\tau_3=+1 (-1)$ for proton (neutron) scattering.
$F_{1,2}^s(Q^2)$ are the strange vector form factors; the strange axial 
vector form factor is commonly denoted as:

\vspace{-0.1in}
\begin{eqnarray}
  F_A^s(Q^2)= \frac{\Delta s}{\left(\displaystyle{1 + {{Q^2}\over{m_A^2} }}
  \right)^2}
\end{eqnarray}

\noindent
where $\Delta s$ is the strange quark contribution to the nucleon spin.

\begin{figure}[h]
  \includegraphics[angle=0,width=16.5pc]{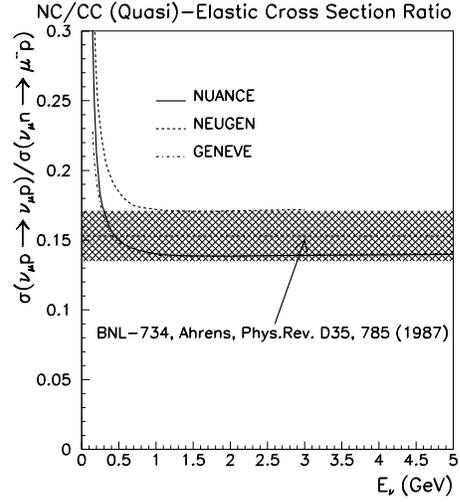}
  \vspace{-0.4in}
  \caption{Most precise measurement of the NC/CC ratio, 
     $(\numu \, p \rightarrow \numu \, p)/(\numu \, n \rightarrow \mu^- \, p)$,
     compared to predictions from NUANCE, NEUGEN, and another neutrino 
     generator, GENEVE~\cite{geneve}.} 
  \label{fig:ncelastic}
\end{figure}

Experiments typically publish NC elastic cross sections with respect to 
the CC QE cross section to minimize systematics. Table~\ref{table:nc-elastic} 
lists a collection of experimental measurements of the NC/CC ratio, 
$(\numu \, p \rightarrow \numu \, p)/(\numu \, n \rightarrow \mu^- \, p)$.
Figure~\ref{fig:ncelastic} shows a comparison between the Monte Carlo
calculations and the most precise measurement of this ratio from BNL 
E734~\cite{nc-elastic-1}. While the BNL E734 result is quoted over a 
particular $Q^2$ range (Table~\ref{table:nc-elastic}), no $Q^2$ restrictions 
have been placed on the Monte Carlo predictions. Also note that NEUGEN 
is $\sim20\%$ larger than the other predictions. This is simply 
because NEUGEN assumes $\Delta s=-0.15$, which enters the differential cross 
section as $(\Delta s)^2$, whereas the other models assume $\Delta s=0$ by 
default~\cite{deltas}.

\vspace{-0.3in}

\vspace{-0.2in}
\begin{table}[h]
\begin{tabular}{|l|c|c|c|}
\hline
Experiment   & Target  & Result  & $Q^2$(GeV$^2$) \\
\hline
BNL E734 \cite{nc-elastic-1} & $CH_2$ & 0.153 $\pm$ 0.018  
                             & $0.5\rightarrow1.0$ \\
BNL CIB  \cite{nc-elastic-2} & $Al$     & 0.11 $\pm$ 0.03   
                             & $0.3\rightarrow0.9$ \\ 
Aachen \cite{nc-elastic-3}   & $Al$     & 0.10 $\pm$ 0.03   
                             & $0.2\rightarrow1.0$ \\
BNL E613 \cite{nc-elastic-4} & $CH_2$   & 0.11 $\pm$ 0.02   
                             & $0.4\rightarrow0.9$ \\
Gargamelle \cite{nc-elastic-5} & $CF_3Br$ & 0.12 $\pm$ 0.06   
                             & $0.3\rightarrow1.0$ \\
\hline
\end{tabular}\\[2pt]
\caption{Several measurements of the ratio,
$(\numu \, p \rightarrow \numu \, p)/(\numu \, n \rightarrow \mu^- \, p)$. 
Also indicated is the $Q^2$ interval over which the ratio was measured.}
\label{table:nc-elastic}
\end{table}

\section{CC and NC Single Pion Production}

The dominant means of single pion production in low energy neutrino 
interactions arises through the excitation of a baryon resonance ($N^*$) 
which then decays to a nucleon-pion final state:

\vspace{-0.2in}
\begin{eqnarray*}
   \numu \, N \rightarrow l &N^*& \\
                         &N^*& \rightarrow \pi \, N^\prime 
\end{eqnarray*}

\noindent
where $N,N^\prime=n,p$. There are seven possible resonant single 
pion reaction channels, three charged current:

\vspace{-0.2in}
\begin{eqnarray*}
   \numu \, p &\rightarrow& \mu^- \, p \, \pi^+ \\
   \numu \, n &\rightarrow& \mu^- \, p \, \pi^0 \\
   \numu \, n &\rightarrow& \mu^- \, n \, \pi^+
\end{eqnarray*}

\noindent
and four neutral current:

\vspace{-0.2in}
\begin{eqnarray*}
   \numu \, p &\rightarrow& \numu \, p \, \pi^0 \\
   \numu \, p &\rightarrow& \numu \, n \, \pi^+ \\
   \numu \, n &\rightarrow& \numu \, n \, \pi^0 \\
   \numu \, n &\rightarrow& \numu \, p \, \pi^- \\
\end{eqnarray*}

\begin{figure}[h]
  \includegraphics[angle=0,width=18.2pc]{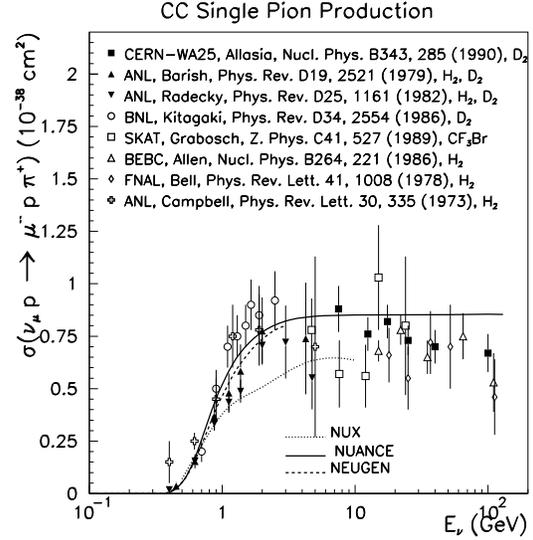}
  \vspace{-0.6in}
  \caption{Measurements of the CC $1\pi$ cross section
           $\sigma(\numu \, p \rightarrow \mu^- p \, \pi^+)$. The Monte Carlos
           assume $m_A=1.032$ GeV and $m_V=0.84$ GeV. The data
           do not include an invariant mass cut with the exception of 
           the CERN-WA25, SKAT, BEBC, and FNAL measurements which are
           reported for $W<2.0$ GeV.}
  \label{fig:cc-pi-1}
\end{figure}

\begin{figure}[h]
  \includegraphics[angle=0,width=17.2pc]{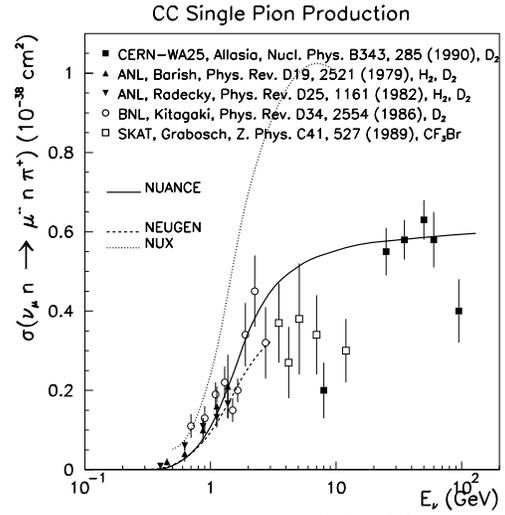}
  \vspace{-0.5in}
  \caption{Measurements of the CC $1\pi$ cross section
           $\sigma(\numu \, n \rightarrow \mu^- n \, \pi^+)$. The data
           do not include an invariant mass cut with the exception of 
           the CERN-WA25 and SKAT measurements which are reported for 
           $W<2.0$ GeV.}
  \label{fig:cc-pi-2}
\end{figure}

\begin{figure}[h]
  \includegraphics[angle=0,width=17pc]{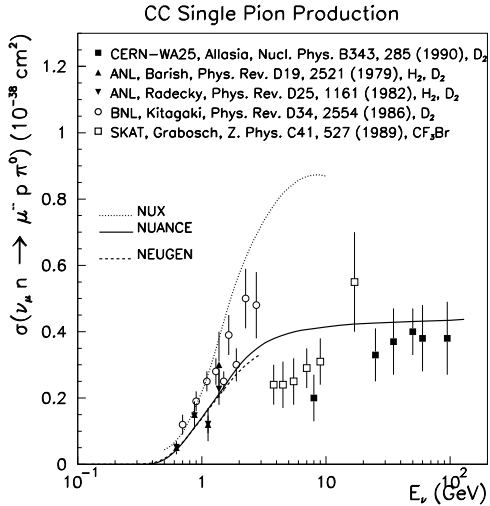}
  \vspace{-0.5in}
  \caption{CC $1\pi$ cross section
           $\sigma(\numu \, n \rightarrow \mu^- p \, \pi^0)$. Same Monte
           Carlo and data criteria as in Figure~\ref{fig:cc-pi-2}.
           Note: the CERN-WA25 data on this channel as 
           reported in the Durham reaction database is actually the
           sum of their measured $\numu \, n \rightarrow \mu^- \, p \, \pi^0$ 
           and $\numu \, n \rightarrow \mu^- \, n \, \pi^+$ cross 
           sections~\cite{cern-wa25}.}
  \label{fig:cc-pi-3}
\end{figure}

\vspace{-0.2in}
\noindent
Traditionally, Monte Carlos base their theoretical calculations of resonant
pion production on the Rein and Sehgal model~\cite{rein-sehgal}. While the 
$\Delta(1232)$ is the dominant resonance at these energies, both the NUANCE 
and NEUGEN generators include additional higher mass resonant states. In 
contrast, the NUX model does not yet contain explicit resonance production, 
and is thus why the NUX predictions exhibit less agreement with the data. 
Such models are commonly tuned to reproduce single pion data, but remain 
poorly constrained because of the limited availibility and large 
uncertainties in this data. With a few exceptions, most of the experimental 
measurements come from deuterium or hydrogen target bubble chamber 
experiments. Figures~\ref{fig:cc-pi-1}-\ref{fig:cc-pi-3} compare the Monte 
Carlo predictions to available CC single pion data. The NUANCE and NEUGEN 
predictions include both resonant and nonresonant contributions and assume no 
invariant mass restrictions. Some caution is warranted in drawing
conclusions from these plots as some of the higher energy data includes a 
$W<2.0$ GeV invariant mass cut. The data sets with $W$ cuts are indicated 
in each caption. 

\vspace{-0.8in} 

\vspace{-1.0in}
\clearpage

\vspace{-0.2in}
The data on NC single pion cross sections is even more limited.
Almost all of this data exists in the form of NC/CC ratios. 
Table~\ref{table:nc-1pi} summarizes the various measurements, all of which 
were conducted using bubble chambers with the exception of BNL and CERN PS 
which both utilized spark chambers. In some instances, the experimental data 
can differ by as much as factors of two or three. Also listed are the 
predictions from NUANCE. In all cases, the NUANCE Monte Carlo agrees with 
at least one measurement.

\begin{table*}[ht]
\newcommand{\m}{\hphantom{$-$}}
\newcommand{\cc}[1]{\multicolumn{1}{c}{#1}}
\begin{tabular}{@{}lllll}
\hline
Source  & Target  & NC/CC Ratio & Value  & Ref \\
\hline
ANL         & $H_2$ 
            & $\sigma(\numu \, p \rightarrow \numu \, p \, \pi^0)/
               \sigma(\numu p \, \rightarrow \mu^- \, p \, \pi^+)$  
            & $0.51 \pm 0.25^*$  & \cite{nc-cc-single-pi-1} \\
ANL         & $H_2$ 
            & $\sigma(\numu \, p \rightarrow \numu \, p \, \pi^0)/
               \sigma(\numu p \, \rightarrow \mu^- \, p \, \pi^+)$  
            & $0.09 \pm 0.05^*$
              & \cite{nc-cc-single-pi-2} \\
NUANCE      & free nucleon
            & $\sigma(\numu \, p \rightarrow \numu \, p \, \pi^0)/
               \sigma(\numu p \, \rightarrow \mu^- \, p \, \pi^+)$  
            & 0.20  & \cite{nuance} \\ 
\hline
ANL         & $H_2$ 
            & $\sigma(\numu \, p \rightarrow \numu \, n \, \pi^+)/
               \sigma(\numu p \, \rightarrow \mu^- \, p \, \pi^+)$  
            & $0.17 \pm 0.08$  & \cite{nc-cc-single-pi-1} \\
ANL         & $H_2$ 
            & $\sigma(\numu \, p \rightarrow \numu \, n \, \pi^+)/
               \sigma(\numu p \, \rightarrow \mu^- \, p \, \pi^+)$  
            & $0.12 \pm 0.04$  & \cite{nc-cc-single-pi-2} \\
NUANCE      & free nucleon
            & $\sigma(\numu \, p \rightarrow \numu \, n \, \pi^+)/
               \sigma(\numu p \, \rightarrow \mu^- \, p \, \pi^+)$  
            & 0.17  &  \cite{nuance} \\
\hline
ANL         & $D_2$ 
            & $\sigma(\numu \, n \rightarrow \numu \, p \, \pi^-)/
               \sigma(\numu n \, \rightarrow \mu^- \, n \, \pi^+)$  
            & $0.38 \pm 0.11$  & 
              \cite{nc-cc-single-pi-3} \\
NUANCE      & free nucleon
            & $\sigma(\numu \, n \rightarrow \numu \, p \, \pi^-)/
               \sigma(\numu n \, \rightarrow \mu^- \, n \, \pi^+)$  
            & 0.27  & \cite{nuance} \\
\hline
Gargamelle  & $C_3H_8$ $CF_3Br$ 
            & $\Sigma_{N=n,p} \: 
               \sigma(\numu \, N \rightarrow \numu \, N \, \pi^0)/
               2\, \sigma(\numu n \, \rightarrow \mu^- \, p \, \pi^0)$  
            & $0.45 \pm 0.08$  & \cite{nc-cc-single-pi-4} \\
CERN PS     & $Al$ 
            & $\Sigma_{N=n,p} \: 
               \sigma(\numu \, N \rightarrow \numu \, N \, \pi^0)/
               2\, \sigma(\numu n \, \rightarrow \mu^- \, p \, \pi^0)$  
            & $0.40 \pm 0.06$  & \cite{nc-cc-single-pi-3} \\
BNL         & $Al$ 
            & $\Sigma_{N=n,p} \: 
               \sigma(\numu \, N \rightarrow \numu \, N \, \pi^0)/
               2\, \sigma(\numu n \, \rightarrow \mu^- \, p \, \pi^0)$  
            & $0.17 \pm 0.04^{**}$  & \cite{nc-cc-single-pi-5} \\
BNL         & $Al$ 
            & $\Sigma_{N=n,p} \: 
               \sigma(\numu \, N \rightarrow \numu \, N \, \pi^0)/
               2\, \sigma(\numu n \, \rightarrow \mu^- \, p \, \pi^0)$  
            & $0.248 \pm 0.085^{**}$ & \cite{nienaber} \\
NUANCE      & free nucleon 
            & $\Sigma_{N=n,p} \: 
               \sigma(\numu \, N \rightarrow \numu \, N \, \pi^0)/
               2\, \sigma(\numu n \, \rightarrow \mu^- \, p \, \pi^0)$  
            & 0.41  & \cite{nuance} \\
\hline
ANL         & $D_2$ 
            & $\sigma(\numu \, n \rightarrow \numu \, p \, \pi^-)/
               \sigma(\numu p \, \rightarrow \mu^- \, p \, \pi^+)$  
            & $0.11 \pm 0.022$  & \cite{nc-cc-single-pi-2} \\
NUANCE      & free nucleon 
            & $\sigma(\numu \, n \rightarrow \numu \, p \, \pi^-)/
               \sigma(\numu p \, \rightarrow \mu^- \, p \, \pi^+)$  
            & $0.19$  & \cite{nuance} \\
\hline
\end{tabular}\\[2pt]
\caption{Measurements of NC/CC single pion cross section ratios.
         The Gargamelle data has been corrected to a free nucleon   
         ratio~\cite{nc-cc-single-pi-4}. Also quoted are the free nucleons 
         cross section predictions from NUANCE assuming $m_A=1.032$ GeV,    
         $m_V=0.84$, and $\stw=0.2319$ in each case. * In their later
         paper~\cite{nc-cc-single-pi-2}, Derrick {\em et al.}
         remark that while this result is $1.6\sigma$ smaller than their
         previous result~\cite{nc-cc-single-pi-1}, the neutron 
         background in this case was better understood. ** The BNL NC 
         $\pi^0$ data was later reanalyzed after properly taking into account
         multi-$\pi$ backgrounds and found to have a larger fractional cross 
         section~\cite{nienaber}.} 
\label{table:nc-1pi}
\end{table*}

\begin{figure}[h]
  \includegraphics[angle=0,width=18.2pc]{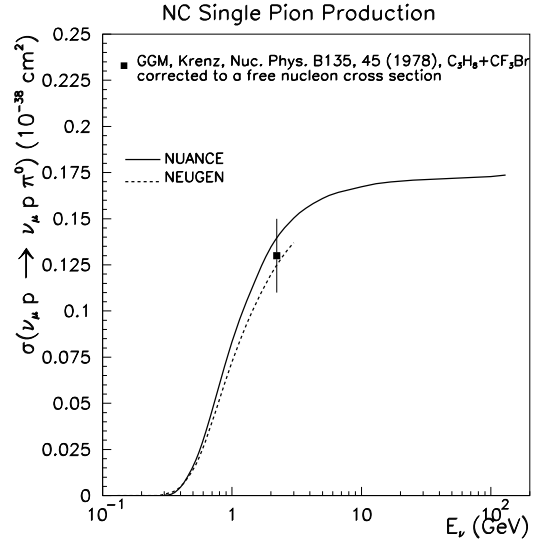}
  \vspace{-0.5in}
  \caption{NC $1\pi$ cross section
           $\sigma(\numu \, p \rightarrow \numu \, p \, \pi^0)$. Shown are the 
           free nucleon cross section predictions from NUANCE and NEUGEN with 
           $m_A=1.032$ GeV, $m_V=0.84$ GeV, and $\stw=0.233$.}
  \label{fig:nc-pi-1}
\end{figure}

\begin{figure}[h]
  \includegraphics[angle=0,width=18.2pc]{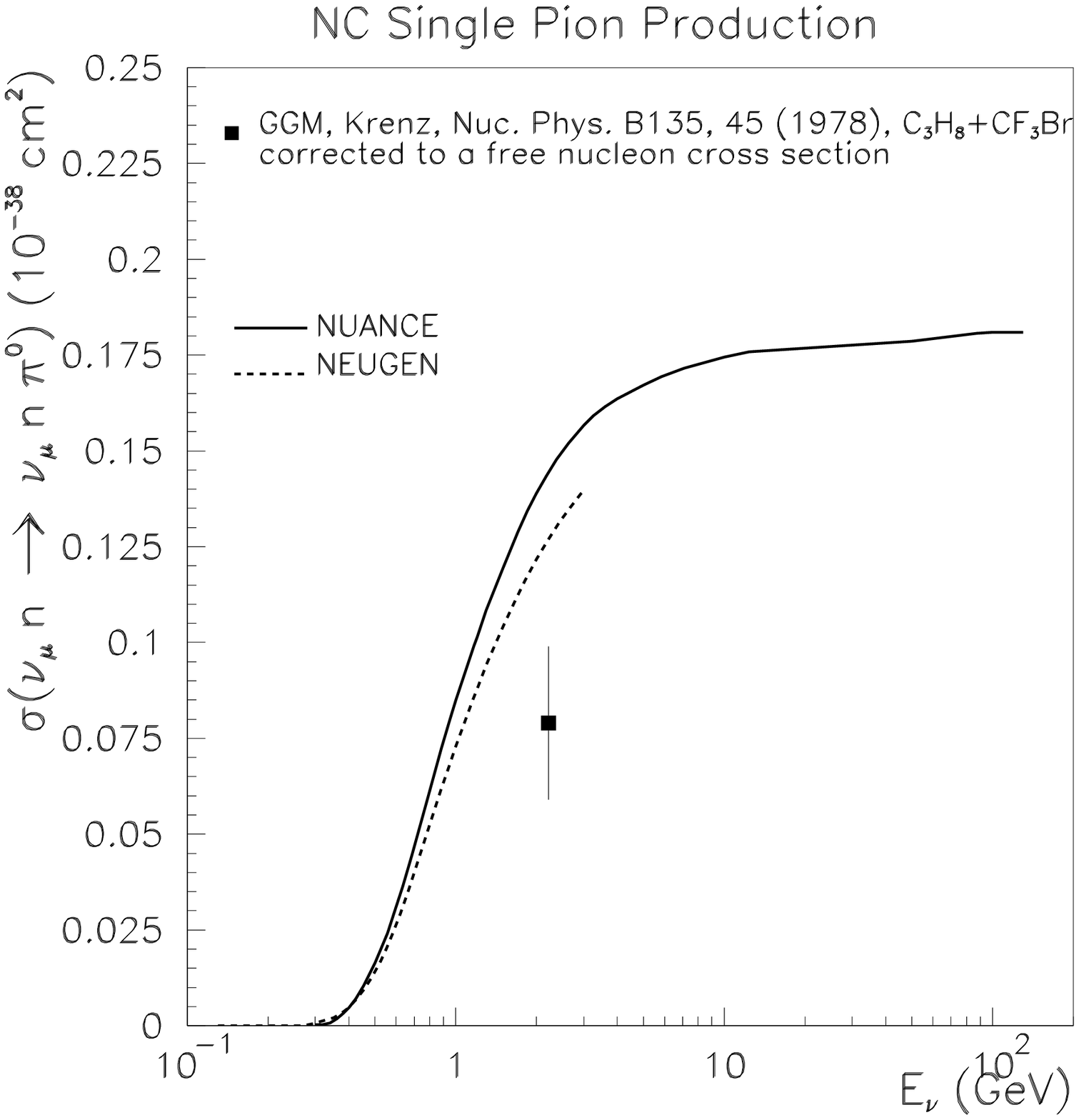}
  \vspace{-0.5in}
  \caption{NC $1\pi$ cross section
           $\sigma(\numu \, n \rightarrow \numu \, n \, \pi^0)$. Same 
           Monte Carlo settings as in Figure~\ref{fig:nc-pi-1}.}
  \label{fig:nc-pi-3}
\end{figure}

\begin{figure}[h]
  \includegraphics[angle=0,width=18.2pc]{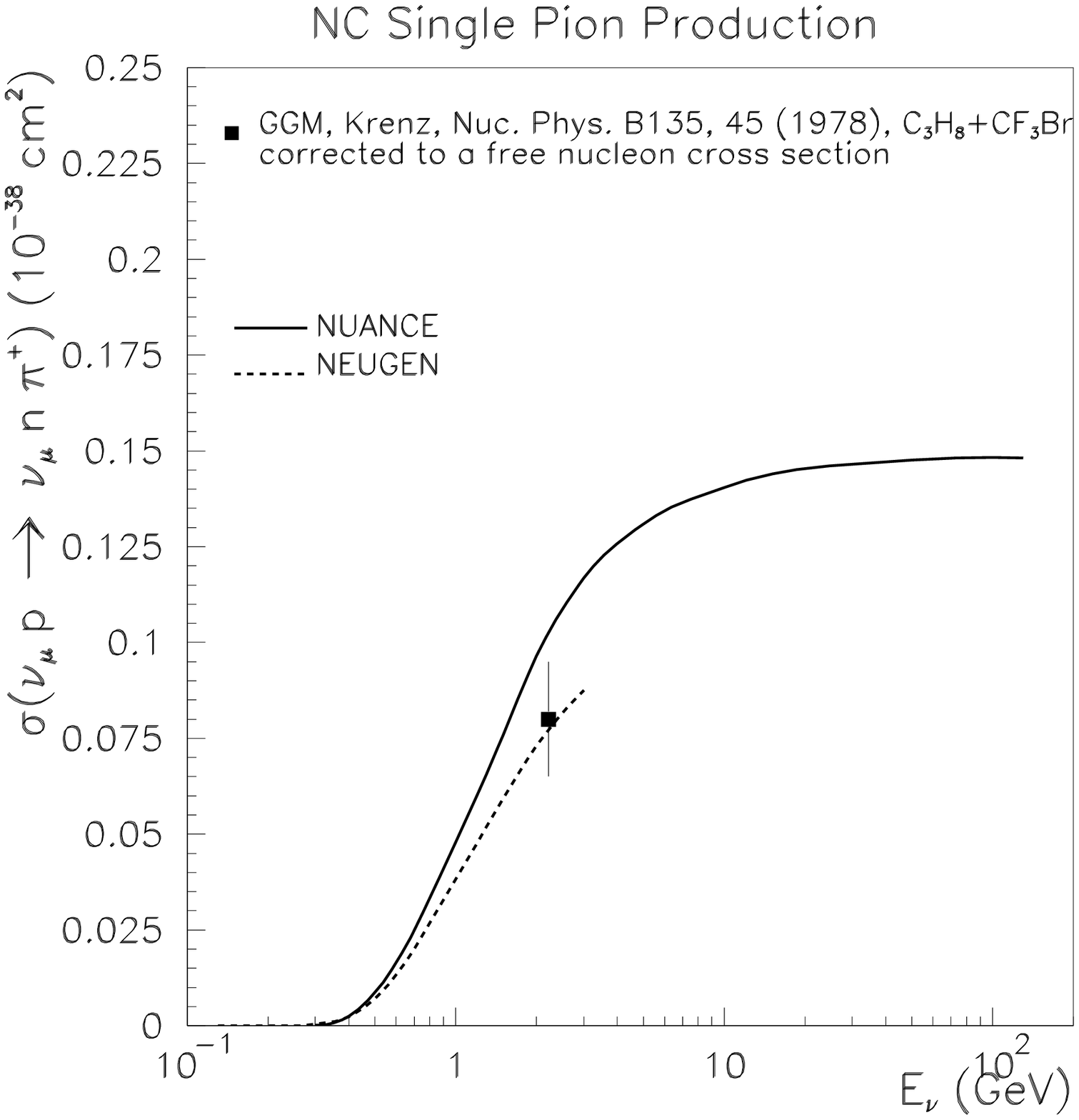}
  \vspace{-0.5in}
  \caption{NC $1\pi$ cross section
           $\sigma(\numu \, p \rightarrow \numu \, n \, \pi^+)$. Same
           Monte Carlo settings as in Figure~\ref{fig:nc-pi-1}.}
  \label{fig:nc-pi-2}
\end{figure}

\begin{figure}[h]
  \includegraphics[angle=0,width=18.2pc]{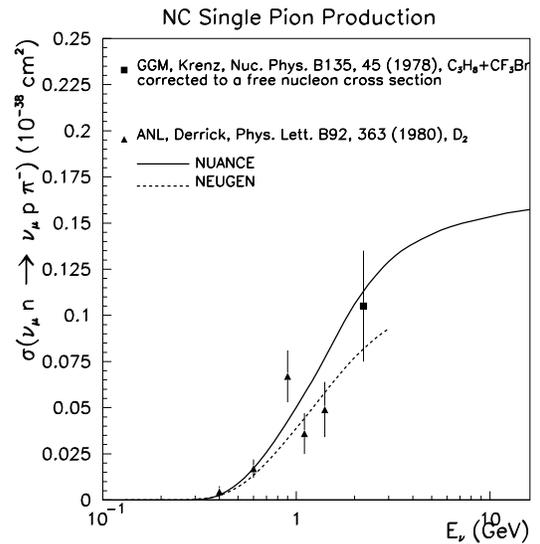}
  \vspace{-0.5in}
  \caption{NC cross section
           $\sigma(\numu \, n \rightarrow \numu \, p \, \pi^-)$. Same
           Monte Carlo settings as in Figure~\ref{fig:nc-pi-1}.}
  \label{fig:nc-pi-4}
\end{figure}

Furthermore, data on {\bf absolute} inclusive NC single pion cross sections is 
extremely sparse. While the ANL 12~ft deuterium bubble chamber 
experiment~\cite{nc-pim}, reported a cross section for the NC $1\pi^-$ 
channel, $\numu \, n \rightarrow \numu \, p \, \pi^-$, the only  
measurements of the remaining NC $1\pi$ cross sections come from a recent 
reanalysis of Gargamelle propane-freon bubble chamber data~\cite{hawker}. 
Figures~\ref{fig:nc-pi-1}-\ref{fig:nc-pi-4} compare these measurements to 
resonant + nonresonant predictions from NUANCE and NEUGEN. Both the data 
and Monte Carlo in all cases have been corrected to free nucleon cross 
sections. 

\vspace{-0.3in}

\clearpage

\section{Single Pion Kinematic Comparisons}

\begin{figure}[h]
  \includegraphics[angle=0,width=20pc]{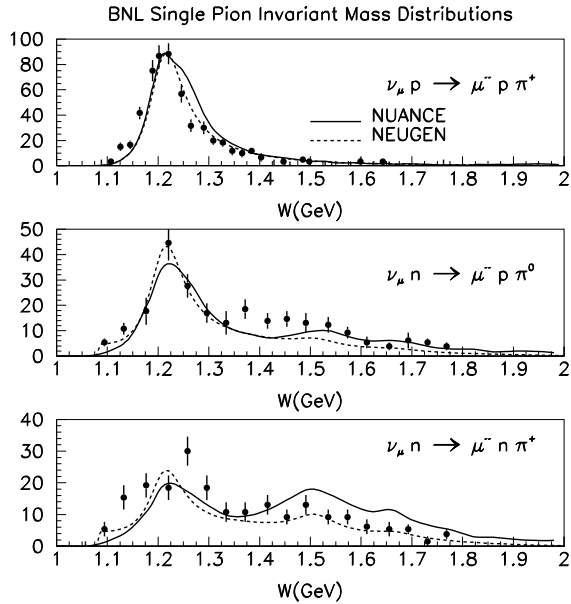}
  \vspace{-0.6in}
  \caption{Invariant mass distributions ($W$) for CC single pion production
           channels as measured in Reference~\cite{bnl-kinematics}. Note
           the clear $\Delta(1232)$ peak.}
  \label{fig:bnl-w}
\end{figure}

\begin{figure}[h]
  \includegraphics[angle=0,width=20pc]{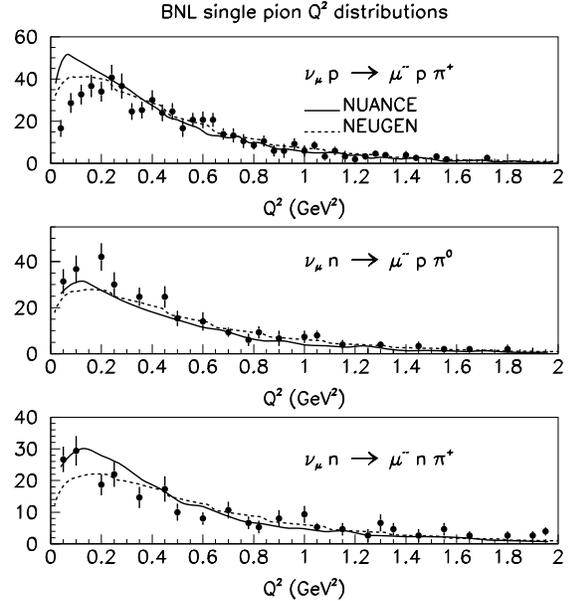}
  \vspace{-0.6in}
  \caption{Four-momentum transfer ($Q^2$) distributions for CC $1\pi$
           channels as measured in~\cite{bnl-kinematics}.}
  \label{fig:bnl-qsq}
\end{figure}

So far these comparisons have been restricted to the case of free nucleon
cross sections; however, comparisons can also be made to measured kinematic 
distributions. Such evaluations were originally performed in testing NEUGEN 
generator performance and presented at NuInt01~\cite{neugen}. The study is
repeated here after including similar NUANCE calculations. 
Figures~\ref{fig:bnl-w}-\ref{fig:bnl-qsq} display invariant mass and $Q^2$ 
distributions for the three CC single pion channels as measured in the 
BNL 7~foot deuterium bubble chamber~\cite{bnl-kinematics}. Both the NUANCE 
and NEUGEN Monte Carlo predictions were generated assuming the BNL flux and 
deuterium target. The Monte Carlo normalization in all plots is determined 
by the peak of the $\numu \, p \rightarrow \mu^- \, p \, \pi^+$ 
invariant mass distribution in the data. Both Monte Carlo models yield 
comparable agreement with the experimental data.

\vspace{-0.2in}

\section{Coherent Pion Production}

In addition to resonance production, neutrinos can also coherently produce
single pion final states. In this case, the neutrino coherently scatters from 
the entire nucleus, transferring negligible energy to the target ($A$). The 
result is a distinctly forward-scattered single pion. Both NC and CC 
coherent pion production processes are possible:

\vspace{-0.2in}
\begin{eqnarray*}
   \numu \, A &\rightarrow& \numu \, A \, \pi^0 \\
   \numu \, A &\rightarrow& \mu^- \, A \, \pi^+ \\
\end{eqnarray*}
\vspace{-0.35in}

\begin{figure}[h]
  \includegraphics[angle=0,width=19pc]{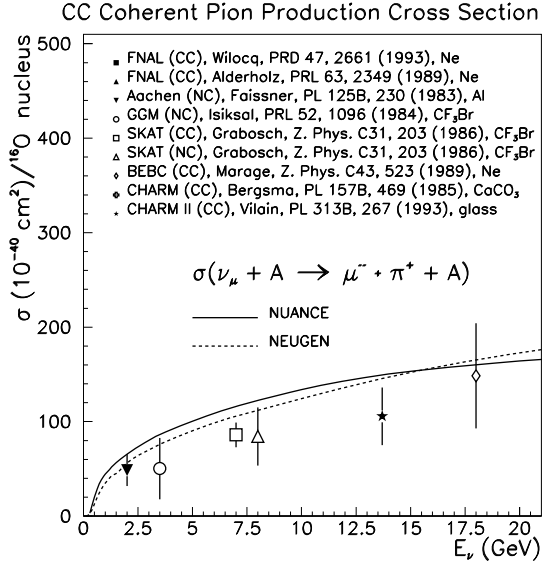}
  \vspace{-0.5in}
  \caption{Coherent pion production data. In this comparison, NUANCE assumes 
           $m_A=1.032$ GeV while NEUGEN uses $m_A=1.0$ GeV.}
  \label{fig:coherent-pi}
\end{figure}

\begin{figure}[h]
  \includegraphics[angle=0,width=19pc]{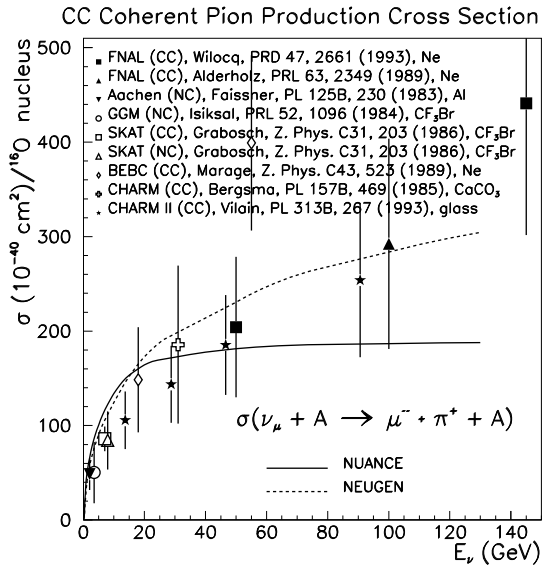}
  \vspace{-0.5in}
  \caption{Same as Figure~\ref{fig:coherent-pi} except extended out to
           higher energies.}
  \label{fig:coherent-pi-alle}
\end{figure}

\noindent
The cross sections for such processes are predicted to be small, but have 
been measured in a variety of neutrino experiments. A comprehensive review of 
the experimental data is provided in Reference~\cite{vilain}. 
Figures~\ref{fig:coherent-pi} and \ref{fig:coherent-pi-alle} show a comparison
of this data to the NUANCE and NEUGEN predictions. Both NC and CC data are 
displayed on the same plot after rescaling the NC data, assuming 
$\sigma^{coh}_{NC} = 1/2 \, \sigma^{coh}_{CC}$~\cite{rein-sehgal-coherent}.
In addition, data from various targets are corrected to oxygen cross sections 
assuming $A^{1/3}$ scaling~\cite{rein-sehgal-coherent}.

\vspace{-0.3in}

\begin{figure}[h]
  \includegraphics[angle=0,width=19pc]{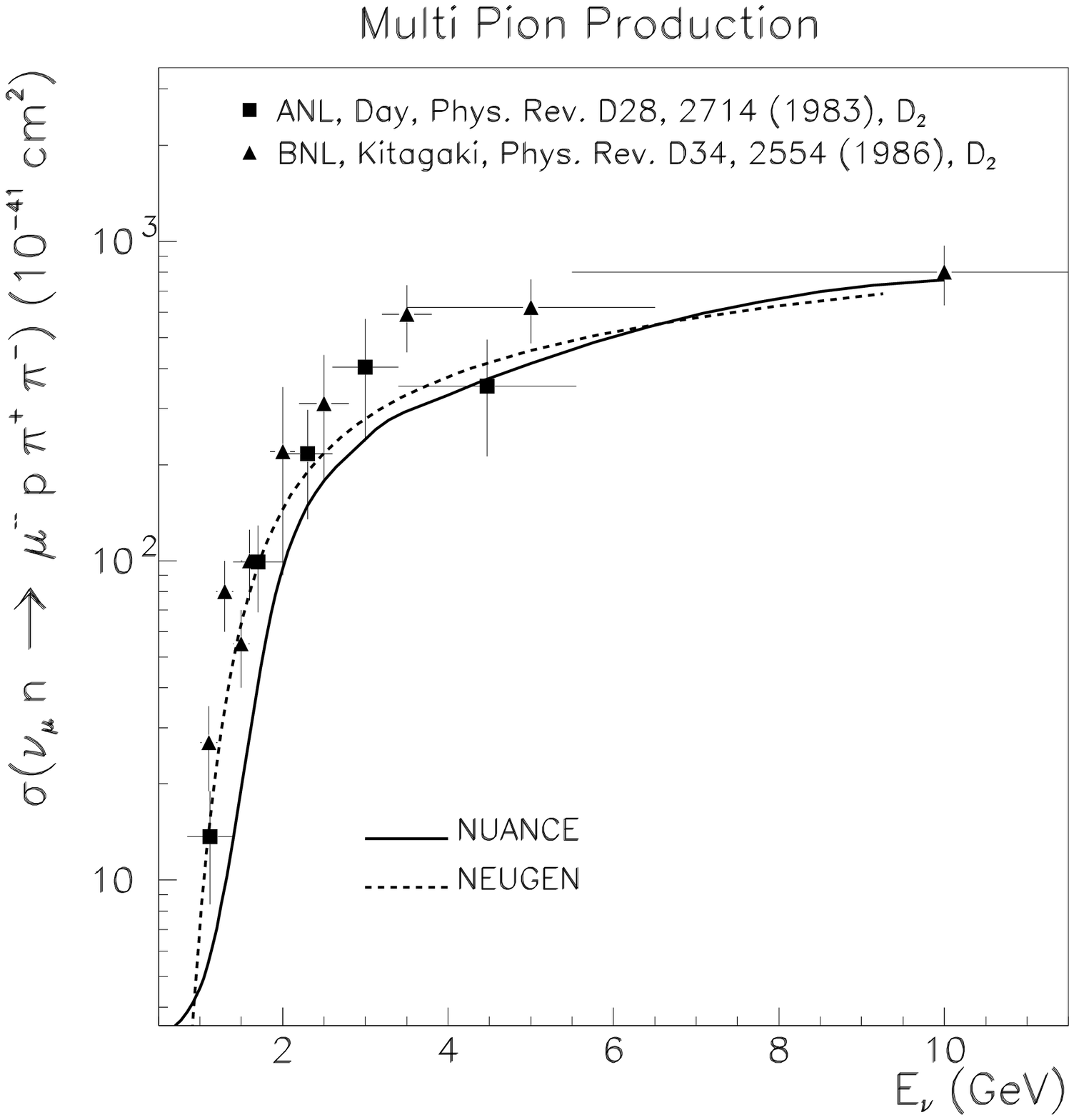}
  \vspace{-0.5in}
  \caption{Cross section for 
           $\numu \, n \rightarrow \mu^- \, p \, \pi^+ \, \pi^-$.}
  \label{fig:multi-pi-1}
\end{figure}

\vspace{-0.3in}
\begin{figure}[h]
  \includegraphics[angle=0,width=19pc]{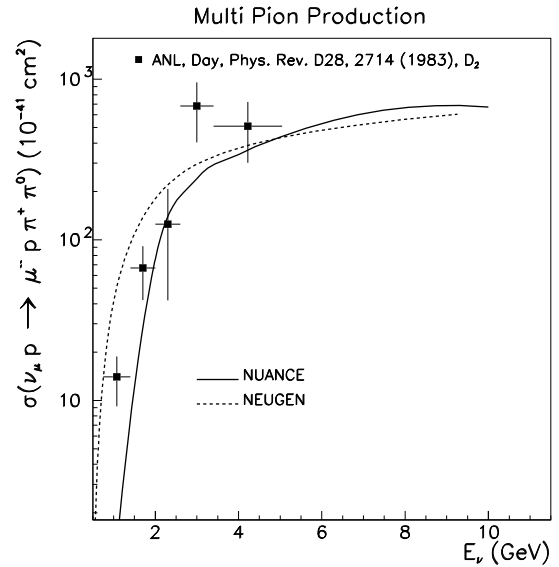}
  \vspace{-0.5in}
  \caption{Cross section for 
           $\numu \, p \rightarrow \mu^- \, p \, \pi^+ \, \pi^0$.}
  \label{fig:multi-pi-2}
\end{figure}
\clearpage

\section{Multi-Pion Production}

\begin{figure}[h]
  \includegraphics[angle=0,width=17pc]{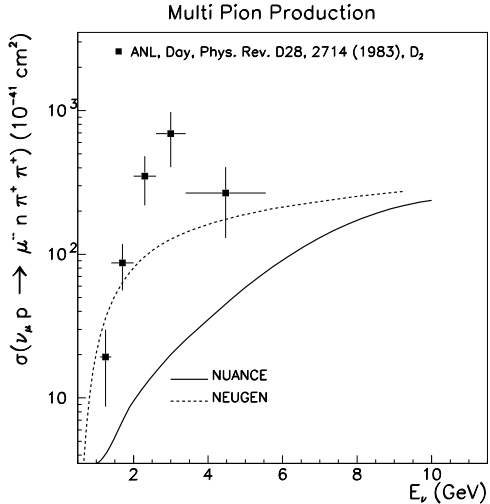}
  \vspace{-0.5in}
  \caption{Cross section for 
           $\numu \, p \rightarrow \mu^- \, n \, \pi^+ \, \pi^+$.}
  \label{fig:multi-pi-3}
\end{figure}

Because of the complexities involved in final state identification 
and isolation, data on multiple pion production is even less 
abundant and less precise than single pion production channels. Likewise, such 
channels exhibit the largest spread in the Monte Carlo estimates. 
Figures~\ref{fig:multi-pi-1}-\ref{fig:multi-pi-3} compare available data 
on dipion production to NUANCE and NEUGEN. Because deep 
inelastic scattering is a large contribution to such cross sections, the 
NUANCE prediction is expected to improve with planned DIS model 
upgrades~\cite{bodek-yang}.

\vspace{-0.3in}
\vspace{-0.3in}

\section{Single Kaon Production}

Proton decay modes containing a final state kaon, $p \rightarrow \nu \, K^+$,
have large branching ratios in many SUSY GUT models. Because there
is a non-zero probability that an atmospheric neutrino interaction can
mimic such a proton decay signature, estimating these background rates is 
an important component to such searches. The following lists some of the 
contributing strange production channels available at low energies:

\vspace{-0.2in}
\begin{eqnarray}
  && \mathrm{CC:} \hspace{1.0in}  \mathrm{NC:} \nonumber \\
  \numu \, n &\rightarrow& \mu^- \, K^+ \, \Lambda^0 \hspace{0.2in}
  \numu \, p \rightarrow \numu \, K^+ \Lambda^0 
\nonumber \\
  \numu \, p &\rightarrow& \mu^- \, K^+ \, p \hspace{0.27in} 
  \numu \, n \rightarrow \numu \, K^0 \, \Lambda^0
\nonumber \\
  \numu \, n &\rightarrow& \mu^- \, K^0 \, p \hspace{0.29in}
  \numu \, p \rightarrow \numu \, K^+ \, \Sigma^0
\nonumber \\\
  \numu \, n &\rightarrow& \mu^- \, K^+ \, n \hspace{0.27in}
  \numu \, p \rightarrow \numu \, K^0 \, \Sigma^+
\nonumber \\
  \numu \, p &\rightarrow& \mu^- \, K^+ \, \Sigma^+ \hspace{0.17in}
  \numu \, n \rightarrow \numu \, K^0 \, \Sigma^0 
\nonumber \\
  \numu \, n &\rightarrow& \mu^- \, K^+ \, \Sigma^0 \hspace{0.18in}
  \numu \, n \rightarrow \numu \, K^+ \, \Sigma^-
\nonumber \\
  \numu \, n &\rightarrow& \mu^- \, K^0 \, \Sigma^+ \hspace{0.17in}
  \numu \, n \rightarrow \numu \, K^- \, \Sigma^+
\end{eqnarray}

\begin{figure}[h]
  \includegraphics[angle=0,width=17.5pc]{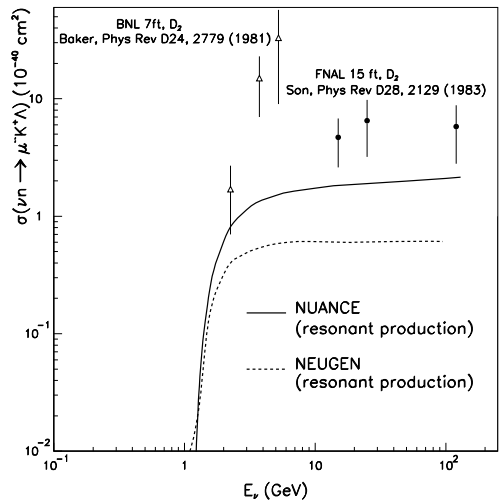}
  \vspace{-0.4in}
  \caption{Measurements of the associated production cross 
           section, $\sigma(\numu \, n \rightarrow \mu^- \, K^+ \, \Lambda^0)$.
           The predictions include resonant contributions only.}
  \label{fig:kaon}
\end{figure}

\noindent
Typically, such reactions have smaller cross sections than their single pion 
counterparts due to the kaon mass and because the kaon channels are not 
enhanced by any dominant resonance (in contrast to $\Delta(1232)$
decays to single pion final states). There are few predictive theoretical 
models for single kaon production~\cite{single-k-models} and little 
experimental data. Most of the data comes from bubble chamber measurements 
where the strange particle decays could be explicitly identified. 
Figure~\ref{fig:kaon} shows the only two experiments which have published 
cross sections on the dominant associated production channel, 
$\numu \, n \rightarrow \mu^- \, K^+ \, \Lambda^0$. Both bubble chamber 
measurements were made on a deuterium target and based on less than 30 events 
combined. To model kaon production, NUANCE and NEUGEN employ the same Rein 
and Sehgal-based framework~\cite{rein-sehgal} as used to model single pion
production, including additional resonance decays other than 
$N^* \rightarrow \pi\, N$. Note that as plotted, neither Monte Carlo includes
DIS contributions to this channel.

\vspace{-0.5in}

\section{Total CC Cross Section}

\begin{figure}[h]
   \includegraphics[angle=0,width=18pc]{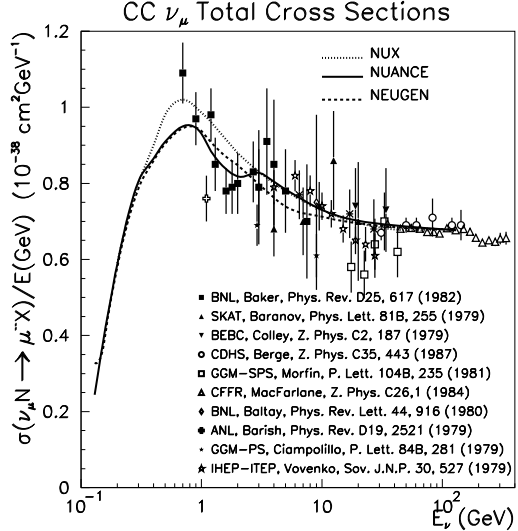}
   \vspace{-0.5in}
   \caption{Total isoscalar inclusive CC cross section per GeV. The 
            predictions are a sum of all CC contributions 
            (i.e., QE, $1\pi$, multi-$\pi$, DIS, etc.).}
   \label{fig:total}
\end{figure}

\begin{figure}[h]
   \includegraphics[angle=0,width=18pc]{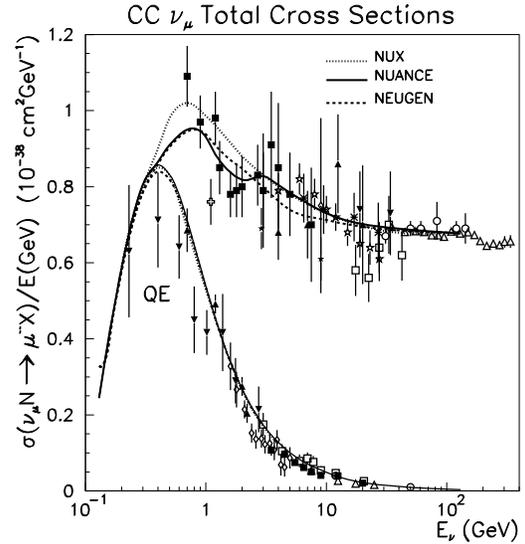}
   \vspace{-0.5in}
   \caption{Same data and Monte Carlo as Figure~\ref{fig:total} with 
            inclusion of QE data from Figure~\ref{fig:qe-data}.}
   \label{fig:total-with-qe}
\end{figure}

Figures~\ref{fig:total}-\ref{fig:total-with-qe} compare total inclusive 
CC cross section predictions to available experimental data. While the total 
cross section at high energy (DIS regime) is known to a few percent, the cross
section at lower energies is much less precisely known. In particular, data 
measurements in the few-GeV range are generally of $\sim10\%$ precision and 
come primarily from experiments which ran in the 1970's and early 
1980's~\cite{naples}. At these energies, it is especially challenging to 
model the total cross section as there are substantial overlapping 
contributions from QE, resonance, and DIS processes. The Monte Carlo 
predictions plotted in Figure~\ref{fig:total} include all of these 
contributions, and are shown to agree fairly well with each other. 

\vspace{-0.2in}

\section{Conclusions}

This work was an attempt to present a comprehensive comparison between
available Monte Carlo generators and experimental data. While the
comparisons are restricted to the case of free nucleon cross sections,
the hope is to expand this study to include neutrino-nucleus cross sections
and predictions from additional Monte Carlo generators. In the comparisons 
shown here, reasonable agreement between the various Monte Carlo simulations 
and experimental data is observed. The agreement is slightly better for CC 
interactions than for NC. However, it is certainly true that the Monte Carlos 
agree with the experimental measurements to the extent that the data agree 
with themselves. Error bars in the current neutrino data set more or less span 
any observed differences in the Monte Carlo predictions. 

\section{Acknowledgments}

It is a pleasure to thank D.~Casper, H.~Gallagher, and P.~Sala for 
their invaluable contributions to this work and for providing their Monte 
Carlo predictions. In addition, the author thanks A.~Bodek, J.~Beacom, 
F.~Cavanna, K.~Datchev, G.~Garvey, E.~Hawker, and O.~Palamara for useful 
input and discussions. Also note that data presented here will be made 
publicly available in the near future~\cite{durham-data}.



\begin{thebibliography}{9}
\bibitem{qe-inputs} $m_V$ ($m_A$) is determined from $e^-$ ($\nu$)
                    scattering data fits, and $g_A$ is precisely determined
                    from neutron beta decay. The best current value for
                    $g_A$ is -1.267; for updated value of $m_A$ see
                    A.~Bodek, these proceedings.
\end{thebibliography}
\end{document}